\documentclass[12pt]{article}
\usepackage{epsf,amsfonts,latexsym,psfrag}
\newsymbol\ltimes 226E
\newsymbol\rtimes 226F
\epsfverbosetrue
\def\double{\mathbb}
\def\ccal{\cal}

\def\cc{{\double C}}

\def\rr{{\double R}}
\def\zz{{\double Z}}
\def\qqq{{\double Q}}
\def\llll{{\double L}}

\def\aa{{\cal A}}
\def\ccc{{\cal C}}
\def\dd{{\cal D}}

\def\hh{{\cal H}}
\def\hhh{{{\double H}}}

\def\mm{{{\ccal M}}}

\def\aa{{\cal A}}
\def\dd{{\cal D}}
\def\hh{{\cal H}}

\def\t{{\rm tr}\,}

\def\ddd{{\,\hbox{$\partial\!\!\!/$}}}

\def\ot{\otimes}
\def\op{\oplus}

\def\bb{\begin{eqnarray}}
\def\ee{\end{eqnarray}}
\def\eee{\nonumber\end{eqnarray}}
\def\pp{\pmatrix}
\def\qq{\quad}

\begin{document}

\hsize 17truecm
\vsize 24truecm
\font\fifteen=cmbx10 at 15pt
\font\twelve=cmbx10 at 12pt
\font\eightrm=cmr8
\baselineskip 18pt

\begin{titlepage}
\setcounter{footnote}{0}
\renewcommand{\thefootnote}{\arabic{footnote}}

\centerline{\twelve CENTRE DE PHYSIQUE TH\'EORIQUE
\footnote{
Unit\'e Propre de Recherche 7061}}
\centerline{\twelve CNRS - Luminy, Case 907}
\centerline{\twelve 13288 Marseille Cedex 9}
\vskip 3truecm

\centerline{\fifteen SPECTRAL ACTION BEYOND THE
STANDARD MODEL}

\vskip 2cm

\begin{center}

   {\bf Thomas SCH\"UCKER}\footnote{\, also at Universit\'e
de Provence, 
\texttt{schucker@cpt.univ-mrs.fr}}\hskip 0.7mm and
{\bf Sami ZOUZOU}\footnote{\, also at Universit\'e de
Constantine,
\texttt{szouzou2000@yahoo.fr}}
\end{center}

\vskip 2cm
\centerline{\bf Abstract}

\medskip

We rehabilitate the
$M_1(\cc)\oplus M_2(\cc)\oplus M_3(\cc)$ model of 
electro-magnetic, weak and strong forces as an almost
commutative geometry in the setting of the spectral action.

\vskip 2truecm
 PACS-92: 11.15 Gauge field theories\\
\indent MSC-91: 81T13 Yang-Mills and other gauge
theories

\vskip 1truecm

\vskip 1truecm
\noindent CPT-01/P.4239\\
\noindent hep-th/yymmxxx

\vskip1truecm

 \end{titlepage}

\section{Introduction}

The associative algebra $\cc\op\hhh\op M_3(\cc)$ allows to
interpret the standard model of electro-magnetic, weak and
strong forces as an almost commutative geometry
\cite{book,tresch}. As such they are naturally unified with
gravity by the spectral action \cite{grav,cc}. In \cite{pris} we
have tried to slightly increase the algebra to $\cc\op
M_2(\cc)\op M_3(\cc)$ in the setting without gravity. This
model suffered from light Higgs scalars generated by
noncommutative geometry. In the present paper we
reconsider the model including gravity.

\section{The fermion masses}

In his approach without gravity, Connes derives the Higgs
field from noncommutative geometry as a fluctuation of a
Yang-Mills connection: the Higgs is an anti-Hermitean 1-form
in a noncommutative differential calculus with exterior
derivative $\delta $ generated by a self-adjoint Dirac operator
$\dd$,
\bb H&=&\sum_j\left\{  a_{0j}\delta a_{1j}
+J\, a_{0j}\delta a_{1j}\,J^{-1}\right\} \cr 
&=&\sum_j \left\{(-i) \rho (a_{0j})[\dd,\rho (a_{1j})]
+J\, (-i)\rho (a_{0j})[\dd,\rho (a_{1j})]\,J^{-1}\right\} ,\ee
where the $a_{0j}$ and $a_{1j}$ are elements of the
associative algebra $\aa$, $\rho $ is a faithful
representation of $\aa$ on the Hilbert space $\hh$ of the
fermions and $J$ is the anti-unitary charge conjugation
operator. The Higgses carry the usual affine {\it group}
representation of connections
\bb ^uH=uHu^{-1}+u\delta u^{-1}
=\rho (u)H\rho (u)^{-1}-i\rho (u)[\dd,\rho (u)^{-1}],\ee
where $u$ is a unitary, $ u\in U(\aa)$.

 We take the algebra
\bb \aa=\cc\op
M_2(\cc)\op M_3(\cc)\,\owns\,(b,a,c).\ee 
The Hilbert
space is copied from the Particle Physics Booklet
\cite{data},
\bb \hh_L&=&
\left(\cc^2\ot\cc^N\ot\cc^3\right)\ \op\ 
\left(\cc^2\ot\cc^N\ot\cc\right), \\
\hh_R&=&\left(\cc\ot\cc^N\ot\cc^3\right)\ 
\op\ \left(\cc\ot\cc^N\ot\cc^3\right)\ 
\op\ \left(\cc\ot\cc^N\ot\cc\right).\ee
 In each summand, the first factor
denotes weak isospin doublets or singlets, the second
denotes
$N$ generations, $N=3$, and the third denotes colour
triplets or singlets.
Let us choose the following basis
of the Hilbert space, counting fermions and
antifermions independently,
$\hh=\hh_L\op\hh_R\op\hh^c_L\op\hh^c_R
=\cc^{90}$: 
\bb
& \pp{u\cr d}_L,\ \pp{c\cr s}_L,\ \pp{t\cr b}_L,\ 
\pp{\nu_e\cr e}_L,\ \pp{\nu_\mu\cr\mu}_L,\ 
\pp{\nu_\tau\cr\tau}_L;&\cr \cr 
&\matrix{u_R,\cr d_R,}\qq \matrix{c_R,\cr s_R,}\qq
\matrix{t_R,\cr b_R,}\qq  e_R,\qq \mu_R,\qq 
\tau_R;&\cr  \cr 
& \pp{u\cr d}^c_L,\ \pp{c\cr s}_L^c,\ 
\pp{t\cr b}_L^c,\ 
\pp{\nu_e\cr e}_L^c,\ \pp{\nu_\mu\cr\mu}_L^c,\ 
\pp{\nu_\tau\cr\tau}_L^c;&\cr\cr  
&\matrix{u_R^c,\cr d_R^c,}\qq 
\matrix{c_R^c,\cr s_R^c,}\qq
\matrix{t_R^c,\cr b_R^c,}\qq  e_R^c,\qq \mu_R^c,\qq 
\tau_R^c.&\eee
This is the current eigenstate basis, the representation
$\rho$ acting on
$\hh$ by
\bb \rho(b,a,c):= 
\pp{\rho_{L}&0&0&0\cr 
0&\rho_{R}&0&0\cr 
0&0&{\bar\rho^c_{L}}&0\cr 
0&0&0&{\bar\rho^c_{R}}}\label{repr0}\ee
with
\bb\rho_{L}(a):=\pp{
a\ot 1_N\ot 1_3&0\cr
0&a\ot 1_N&},\qq
\rho_{R}(b):= \pp{
b 1_N\ot 1_3&0&0\cr 0&\bar b 1_N\ot 1_3&0\cr 
0&0&\bar
b1_N}, \label{repr1}
\ee\bb 
  \rho^c_{L}(b,c):=\pp{
1_2\ot 1_N\ot c&0\cr
0&\bar b1_2\ot 1_N},\qq
\rho^c_{R}(b,c) := \pp{
1_N\ot c&0&0\cr 0&1_N\ot c&0\cr
0&0&\bar b1_N}.  \label{repr2} \ee
 The
apparent asymmetry between particles and
antiparticles -- the former are subject to weak, the
latter to strong interactions -- disappears after
application of the spin lift which involves the 
charge conjugation
\bb J=\pp{0&1_{15N}\cr 1_{15N}&0}\circ 
\ {\rm complex\ conjugation}.\ee
 For the sake of
completeness, we record the chirality as matrix
\bb \chi=\pp{-1_{8N}&0&0&0\cr 0&1_{7N}&0&0\cr
 0&0&-1_{8N}&0\cr 0&0&0&1_{7N} }.\ee
The Dirac operator
\bb \dd=\pp{0&\mm&0&0\cr 
\mm^*&0&0&0\cr 
0&0&0&\bar\mm\cr 
0&0&\bar\mm^*&0}\ee
is constructed from the fermionic mass matrix of the standard
model,
\bb\mm=\pp{
\pp{1&0\cr 0&0}\ot M_u\ot 1_3\,+\,
\pp{0&0\cr 0&1}\ot M_d\ot 1_3
&0\cr
0&\pp{0\cr 1}\ot M_e},\ee
with
\bb M_u:=\pp{
m_u&0&0\cr
0&m_c&0\cr
0&0&m_t},&& M_d:= C_{KM}\pp{
m_d&0&0\cr
0&m_s&0\cr
0&0&m_b},\\[2mm] && M_e:=\pp{
m_e&0&0\cr
0&m_\mu&0\cr
0&0&m_\tau},\ee
with $C_{KM}$ denoting the Cabibbo-Kobayashi-Maskawa
matrix. 

The intersection form
\bb \cap_{ij}:=\t
\left[\chi\,\rho(p_i)\,J\rho(p_j)J^{-1}
\right],\ee 
with a set of
minimal projectors $p_j$ in $\aa$ is non-degenerate and
Poincar\'e duality holds. Indeed, we have
$p_1=\left(1,0,0\right) $, $p_2=\left(0,\pp{1&0\cr
0&0},0\right)$,
$p_3=\left(0,0,\pp{1&0&0\cr 0&0&0\cr 0&0&0}\right) $, and
\bb \cap=N\pp{2&-1&2\cr -1&0&-1\cr 2&-1&0}.\ee
Note that for Dirac instead of Weyl neutrinos in all three
generations, $\hh=\cc^{96}$, Poincar\'e duality would fail:
\bb \cap=N\pp{4&-1&2\cr -1&0&-1\cr 2&-1&0}.\ee
Since Majorana masses are excluded in Connes'
noncommutative geometry, we arrive at the same conclusion
as in the $\cc\op\hhh\op M_3(\cc)$ version: Poincar\'e
duality allows at most two out of three massive neutrinos. For
completeness we recall the intersection form of the
quaternionic version with Weyl or Dirac neutrinos:
\bb N\pp{2&-2&2\cr -2&0&-2\cr 2&-2&0}\qq{\rm or}\qq
N\pp{4&-2&2\cr -2&0&-2\cr 2&-2&0}.\ee

Let us come back to our example and compute the Higgs
representation. It is made of two isospin doublets, colour
singlets $h_1$ and
$h_2$,
\bb H&=&  i\pmatrix
{0&\rho_L(h)\mm&0&0\cr \mm^*\rho_L(h^*)&0&0&0
\cr 0&0&0&\overline{\rho_L(h)\mm}\cr
0&0&\overline{\mm^*\rho_L(h^*)}&0},\\&&
h=\pp{h_1&h_2}=
\pp{h_{11}& h_{12}\cr  h_{21}&  h_{22}}
\in
M_2(\cc).\label{higgs}\ee
The Higgs being a fluctuation of the Yang-Mills connection,
the Higgs Lagrangian is derived from the Yang-Mills
Lagrangian in noncommutative geometry \cite{book,tresch}.
This computation leads to the three pieces, that are added by
hand in the conventional theory: the kinetic term with its
minimal coupling to the Yang-Mills fields, the Higgs potential
with its spontaneous symmetry breaking and the Yukawa
couplings to the fermions. The ground state of the Higgs
potential is
$H_0=0$ and in terms of the scalar
variable $\Phi =H-i\dd$, that transforms {\it homogeneously},
$ ^u\Phi =\rho (u)\Phi \rho (u)^{-1}$, the ground state, that
gives masses to the fermions, is the Dirac operator
$\Phi _0=-i\dd$. This justifies the identification of Dirac
operator and fermionic mass matrix above. The ground state
also gives masses to the physical scalars and the above choice
has neutral and charged ones, that are too light to withstand
confrontation with experiment. 

In his second approach, Connes uses the same
noncommutative geometry to derive Yang-Mills connections
and their Higgs scalars as fluctuations of the gravitational
field \cite{grav}. Now the Higgs is computed as:
\bb H=\sum_j\left\{  \rho (u_{j})[\dd,\rho (u_{j})^{-1}]
+J\, \rho (u_{j})[\dd,\rho (u_{j})^{-1}]\,J^{-1}\right\}
,\qq  u_j\in U(\aa).\label{spectralhiggs}\ee
In our example, $U(\aa)=U(1)\times U(2)\times U(3)$, we get
the same result as without gravity: two isospin doublets of
scalars,
$\varphi _1$ and $
\varphi _2$ in the homogeneous notation. 

The Higgs being a fluctuation of the metric, the Higgs
Lagrangian is now derived from the 
noncommutative version of general relativity, the spectral
action
\cite{cc}. This computation leads again to the three pieces,
 kinetic
term with its minimal coupling,
Higgs potential and
 Yukawa couplings. The Higgs potential breaks again the
gauged group $U(\aa)$ spontaneously. Here however, it is
not true in general that the ground state is given by the Dirac
operator, $\Phi _0=-i\dd$. This holds true for the standard
model with algebra
$\aa=\cc\op\hhh\op M_3(\cc)$. The first counter example
with
$\Phi _0 \not=-i\dd$
is due to Girelli
\cite{florian} with $\aa=\hhh\op\hhh$. We will show that
replacing the quaternions $\hhh$ by complex $2\times 2$
matrices in the standard model also spoils this precious
property. We say precious because it allows for different
masses within the same irreducible fermion multiplet.

In our example, writing the two complex doublets $\varphi
_1,\ \varphi _2$ as a complex $2\times 2$ matrix $\varphi $
like we did before with its inhomogeneous counter part $h$
in equation (\ref{higgs}),  the Higgs potential takes the simple
form,
\bb V(\varphi )=
\lambda \t [(\Phi ^*\Phi)^2 ]\,-\,{\textstyle\frac{1}{2}}
\mu ^2\t[\Phi ^*\Phi] =4\lambda \t [(\mm\mm^*\rho
_L(\varphi ^*\varphi ))^2]\,-\,2\mu ^2\t[\mm\mm^*\rho
_L(\varphi ^*\varphi )],
\ee
with positive parameters $\lambda $ and $\mu $. For
simplicity let us put all fermion masses to zero except for the
top and bottom mass that we take different and let us put
the Cabibbo-Kobayashi-Maskawa matrix to one. Then the
ground state is
\bb \varphi _0=\,\frac{\mu }{2\sqrt \lambda }\, \pp{
m_t^{-1}&0\cr 0&m_b^{-1}}.\ee
But this means that the spontaneous symmetry breaking
induces identical top and bottom masses and the input
values $m_t$ and $m_b$ in the initial Dirac operator have
nothing to do with fermion masses. 

\section{Spin lifts with central extensions}

In the spectral action setting our model has three
shortcomings, $m_t=m_b$, light physical scalars and two
additional $U(1)$ bosons, which are 
anomalous and also too light. In this section we centrally
extend the spin lift, that is necessary to fluctuate the
gravitational field. We motivate these extensions by imposing
the lift to be double-valued and show that the most
economical such extension solves all three shortcomings.
Indeed it recuperates the standard model as Connes derives it
from the algebra 
$\cc\op\hhh\op M_3(\cc)$.

\subsection{Commutative geometry}

To fluctuate the gravitational field Connes proceeds in two
steps. In the first step, he reformulates Einstein's derivation of
general relativity from Riemannian geometry in the algebraic
language of his geometry by carefully avoiding to use the
commutativity of the underlying algebra
$\aa=\ccc^\infty(M)$ of differentiable functions on spacetime
$M$. Here the key ingredient is a group homomorphism $L$
that maps every general coordinate transformation $\varphi $
of spacetime to a gauged Lorentz transformation acting on
Dirac spinors. This homomorphism generalizes the spin lift
$SO(3)\rightarrow SU(2)$ of quantum mechanics to the
special and general relativistic setting. The local form of this
double-valued homomorphism is spelled out in \cite{lift}. The
algebraic formulation of coordinate transformations is
the group of automorphisms of the algebra, Aut$(\aa)$. 
Note that although
$\aa=\ccc^\infty(M)$ is commutative, its automorphism
group, Aut$(\aa)$=Diff$(M)$ is highly nonAbelian. The
algebraic definition of the gauged spin group is what
Connes calls the group of automorphisms lifted to the Hilbert
space 
\bb{\rm Aut}_\hh(\aa):=\{U\in {\rm End}(\hh),\
UU^* =U^*U=1,\ UJ=J U,\ U\chi =\chi U,\ 
i_U\in{\rm Aut}(\rho (\aa))\},\ee
with $i_U(x):=UxU^{-1}$. In Riemannian geometry the Hilbert
space consists of square integrable Dirac spinors, the chirality
is $\chi  =\gamma _5$ and $J$ is the charge conjugation of
Dirac spinors.
 The first three properties
say that a lifted automorphism $U$ preserves probability,
charge conjugation and chirality. The
fourth, called {\it covariance property} is related to the
locality requirement of field theory. It allows to define the
projection
$p:\ {\rm Aut}_\hh(\aa)\longrightarrow {\rm
Aut}(\aa)$ by
\bb p(U)=\rho ^{-1}i_U\rho \ee
Of course we demand that the lift respects the projection,
$p(L(\varphi ))=\varphi$. 

 Einstein shows
that the gravitational field is coded in a Riemannian metric
and Connes shows that the Riemannian metric is coded in its
Dirac operator $\ddd$. Furthermore starting from the flat
Dirac operator $\ddd_0$ one gets a curved one
$\ddd=L(\varphi )\ddd_0L(\varphi )^{-1}$ by fluctuating
with a general coordinate transformation $\varphi $. The
spectral action is simply the trace of $\ddd$ properly
regularized and it reproduces the Einstein-Hilbert action.
Centrally extending $L$ to the group of unitaries
$U(\aa)=\,^MU(1)$ yields gravity coupled to Maxwell's
electromagnetism. However welcome, this central extension is
optional in the sense that it does not change the degree of
valuedness of the lift $L$. 

In the second step, Connes repeats his derivation of general
relativity for almost commutative geometries, that is tensor
products of the infinite dimensional, commutative algebra
$\ccc^\infty(M)$ with finite dimensional, noncommutative
algebras. It is in this precise context that he derives some very
special Yang-Mills-Higgs models by fluctuating the metric. 

\subsection{Central extensions of finite geometries}

We may restrict ourselves to the finite dimensional, `internal'
part where central extensions of the spin lift are readily
available \cite{serge}. Let
$\aa$ be a real, associative involution algebra with unit, that
admits a faithful  * representation $\rho $. In
finite dimensions, a simple such algebra is a real, complex or
quaternion matrix algebra,
$\aa= M_n(\rr),\ M_n(\cc)$ or
$M_n(\hhh) $, represented irreducibly on the Hilbert
space
$\hh=
\rr^n,\ \cc^n$ or $\cc^{2n}$. In the first and third
case, the representations are the fundamental ones,
$\rho(a)=a,\ a\in\aa$, while $M_n(\cc)$ has two
non-equivalent irreducible representations on
$\cc^n$, the fundamental one, $\rho(a)=a$ and its
complex conjugate $\rho(a)=\bar a$. In the general
case we have sums of simple algebras and sums of
irreducible representations as the model under consideration.
With this application in mind, we concentrate on complex
matrix algebras
$M_n(\cc)$ in this section. Anyhow, $M_n(\rr)$
and
$M_n(\hhh)$,  do not have central unitaries close to the
identity. In the following it will be important to separate the
commutative and noncommutative parts of the algebra:
\bb\aa=\cc^M\oplus
\bigoplus_{k=1}^N M_{n_k}(\cc)\ \owns
a=(b_1,...b_M,c_1,...,c_N),\qq n_k\geq 2.
\label{algebra}\ee
Its group of unitaries is
\bb U(\aa)=U(1)^M\times
\matrix{N\cr \times\cr {k=1}} U(n_k)\
\owns\ u=(v_1,...,v_M,w_1,...,w_N)\ee
and its group of central
unitaries 
\bb U^c(\aa):=U(\aa)\cap\,{\rm
center}(\aa)=U(1)^{N+M}\
\owns\ u_c=
( v_{c1},...,v_{cM},w_{c1}1_{n_1},...,w_{cN}1_{n_N}).\ee
The component of the automorphism group ${\rm
Aut}(\aa)$, that is
connected to the identity, is the group of inner
automorphisms,
${\rm Aut}(\aa)^e={\rm In}(\aa)$. There are additional,
discrete automorphisms, the complex conjugation and, if
there are identical summands in $\aa$, their permutations.
These discrete automorphisms do not concern us here. An
inner automorphism is of the form $i_u(a)=u a u^{-1}$ for
some unitary $ u \in U(\aa)$. 
Multiplying $u$ with a central unitary
$u_c$ of course does not affect the inner automorphism
$i_{u_cu}=i_u$. Note that this ambiguity distinguishes
between `harmless' central unitaries $v_{c1},...,v_{cM}$ and
the others, $w_{c1},...,w_{cN}$, in the sense that
\bb {\rm In}(\aa)=U^n(\aa)/U^{nc}(\aa),\label{inntrue}\ee
where we have defined the group of noncommutative unitaries
\bb U^n(\aa):=\matrix{N\cr \times\cr {k=1}} U(n_k)\
\owns\ w\ee
and $U^{nc}(\aa):=U^n(\aa)\cap U^c(\aa) \owns w_c$.
The map 
\bb i:U^n(\aa)&\longrightarrow&{\rm In}(\aa)\cr 
(1,w)&\longmapsto&i_w\ee
 has kernel Ker$\,i=U^{nc}(\aa)$. 

The lift of an inner
automorphism to the Hilbert space has a natural
form
\cite{tresch}, $L=\hat L\circ i^{-1}$ with
\bb \hat L(w)=\rho(1,w)J\rho(1,w)J^{-1}.\ee
It satisfies $p(\hat L(w))=i(w)$.
If the kernel of $i$ is contained in the kernel of $\hat L$ then
the lift is well defined, as e.g. for $\aa=\hhh$,
$U^{nc}(\hhh)=\zz_2$. 
\begin{eqnarray}
&&{\rm Aut}_\hh(\aa)\nonumber \\
&& \hskip -2mm p\ 
\parbox{6mm}{\begin{picture}(20,10)
\put(0,15){\vector(0,-1){30}}
\put(15,-15){\vector(-1,4){8}}
\put(15,-15){\vector(0,1){33}}
\end{picture}}
 L 
\parbox{8mm}{\begin{picture}(20,10)
\put(30,-15){\vector(-1,2){16}}
\put(32,-10){$\hat{L}$}
\end{picture}}
\parbox{12mm}{\begin{picture}(20,10)
\put(65,-15){\vector(-2,1){67}}
\end{picture}}
\ell
\\[4mm]
&&{\rm In}(\aa)\stackrel{i}{\longleftarrow} 
U^n(\aa)\begin{array}{c}\\[-3mm]
\hookleftarrow \\[-5mm]
\stackrel{\vector(1,0){15}}{\mbox{\footnotesize $\det$}} 
\end{array}
U^{nc}(\aa) \nonumber
\end{eqnarray}
For more complicated real or
quaternionic algebras, $U^{nc}(\aa)$ is finite and the lift $L$
is multi-valued with a finite number of values. For
noncommutative, complex algebras, their continuous family of
central unitaries can not be eliminated except for very special
representations and we face a continuous infinity of values.
The solution of this problem is to extend $\hat L$ by the
harmful central unitaries $w_c\in U^{nc}(\aa)$:
\bb \ell(w_c)&=&\rho\!\! \left(
\prod_{j_1=1}^N(w_{cj_1})^{q_{1j_1}},
...,\prod_{j_M=1}^N(w_{cj_M})^{q_{Mj_M}},\right.&\cr &&
\left.\qq
\prod_{j_{M+1}=1}^N(w_{cj_{M+1}})^{q_{{M+1},j_{M+1}}}
1_{n_1},
...,\prod_{j_{M+N}=1}^N(w_{cj_{M+N}})^{q_{{M+N},j_N}}
1_{n_N}
\right)
 J
\rho (...)
\,J^{-1}
\label{ell}\ee
with the $(M+N)\times N$ matrix of charges $q_{kj}$, charge
because in the commutative case there is only one
(harmless) $U(1)$ and $q$ is the electric charge. We allow
multi-valued group homomorphisms, $q_{kj}\in\qqq$. The
general extension satisfies indeed
$p(\ell(w_c))=1\in{\rm In}(\aa)$ for all $w_c\in
U^{nc}(\aa)$. 

Having adjoined the harmful, continuous central unitaries, we
may now stream line our notations and write the group of
inner automorphisms as 
\bb {\rm In}(\aa)=\left( 
\matrix{{N} \cr \times \cr k=1} SU(n_k)\right) /\Gamma 
\owns[w_\varphi] = [(w_{\varphi 1},...,w_{\varphi N})]\ {\rm
mod}\
\gamma \label{innfake} .\ee
$\Gamma $ is the discrete group
\bb\Gamma=\matrix{{N}\cr \times\cr
{k=1}}\zz_{n_k}\ \owns\ (z_11_{n_1},...,z_N1_{n_N}),\qq
z_{k}=\exp[-m_{k}2\pi i/n_k],\ m_k=0,...,n_k-1
. \label{discrete}\ee
The quotient is factor by factor. This way to write inner
automorphisms is convenient for complex matrices, but not
available for real and quaternionic matrices. Equation
(\ref{inntrue}) remains the general characterization of
inner automorphisms.

The lift $L(w_\varphi )=(\hat L\circ i^{-1})(w_\varphi )$
is multi-valued with, depending
on the representation, up to $ |\Gamma |=\prod_{j=1}^N n_j$
values. More precisely the multi-valuedness of $L$ is indexed
by the elements of the kernel of the projection $p$ restricted
to the image $L({\rm In}(\aa))$. Depending on the choice of
the charge matrix
$q$, the central extension $\ell$ may reduce this
multi-valuedness. Extending harmless central unitaries is
useless for any reduction. With
the multi-valued group homomorphism
\bb (h_\varphi,h_c) : U^n(\aa)&\longrightarrow & {\rm
In}(\aa)\times U^{nc}(\aa)\cr  
(w_j) & \longmapsto &((w_{\varphi j} , w_{cj}))=((w_j(\det
w_j)^{-1/n_j},(\det w_j)^{1/n_j}))\label{isom},\ee
 we can write the two lifts
$L$ and
$\ell$  
together in closed form
$\llll:U^n(\aa)\rightarrow
{\rm Aut}_\hh(\aa)$:
\bb\llll(w)&=&L(h_\varphi (w))\,\ell(h_c(w))\cr \cr 
&=&
\rho\!\! \left(
\prod_{j_1=1}^N(\det w_{j_1})^{\tilde q_{1j_1}},
...,\prod_{j_M=1}^N(\det w_{j_M})^{\tilde
q_{Mj_M}},\right.\cr &&\left.\qq
w_1\prod_{j_{M+1}=1}^N(\det w_{j_{M+1}})^{\tilde
q_{{M+1},j_{M+1}}}, ...,w_N\prod_{j_{N+M}=1}^N(\det
w_{j_{N+M}})^{\tilde q_{{N+M},j_{N+M}}}\right)
\nonumber\\[2mm]
&&\times\,
J \rho (...) J^{-1}.\label{generallift}\ee
We have set 
\bb\tilde q:=\left( q-\pp{0_{M\times N}\cr\cr  1_{N\times
N}}
\right) \pp{n_1&&\cr &\ddots&\cr &&n_N}^{-1}.\ee
Due to the
phase ambiguities in the roots of the determinants, the
extended lift
 $\llll$ is multi-valued in general. We will impose it to be
double-valued because this is the valuedness of the lift in the
commutative case, extended or not. In addition we will impose
the extended lift to be even,
$\llll (-u)=\llll (u)$,  which translates into
conditions on the charges, conditions that depend on the
details of the representation $\rho $. This property is
motivated from the observation that the conjugation $i$ is
even. Also, in the case where the associative algebra $\aa$ has
no commutative part $\cc$, and consequently no harmless
$U(1)$, the homomorphism $\hat L$ is even.

\subsection{The $M_1(\cc)\oplus M_2(\cc)\oplus M_3(\cc)$
example}

The algebra has two harmful $U(1)$s and the representation
(\ref{repr0}-\ref{repr2}) yields the extended lift,
\bb \llll(w_{1},w_{2})=\mathrm{diag}\pmatrix{ w_1\ot
1_N\ot w_2\ (\det w_1)^{\tilde q_{21}+\tilde q_{31}} (\det
w_2)^{\tilde q_{22}+\tilde q_{32}}\cr w_1\ot 1_N\ (\det
w_1)^{-\tilde q_{11}+\tilde q_{21}} (\det w_2)^{-\tilde
q_{12}+\tilde q_{22}}\cr 1_N\ot w_2\ (\det w_1)^{\tilde
q_{11}+\tilde q_{31}} (\det w_2)^{\tilde q_{12}+\tilde
q_{32}}\cr 1_N\ot w_2\ (\det w_1)^{-\tilde q_{11}+\tilde
q_{31}} (\det w_2)^{-\tilde q_{12}+\tilde q_{32}}\cr 1_N\ (\det
w_1)^{-2\tilde q_{11}} (\det w_2)^{-2\tilde q_{12}}\cr {\rm
complex\ conjugate}},\ee 
with $w_1\in U(2)$, $w_2\in U(3)$.
This lift is even in the following two cases: $\tilde q_{12}$ 
even and both
$\tilde q_{22}$ and $\tilde q_{32}$ odd or  
$\tilde q_{12}$ is odd and both
$\tilde q_{22}$ and $\tilde q_{32}$ are even. For the lift to
be double-valued  the entries of the first column of $\tilde q$
must be quarter integers. More precisely the most general
double-valued and even lift has a charge matrix
\bb 	\tilde q=\pp{
z_1+k_1/4&2z_4+1+z_5\cr 
z_2+k_2/4&z_5\cr 
z_3+k_3/4&2z_6+z_5},\ee
with six integers $z_j$. The three $k_j$ are either from the
set $\{0,2\}$ or from $\{1,3\}$, the three $k_j$ cannot be
all 0 and they cannot simultaneously take the value 2.
The corresponding matrix $q$ is,
\bb 	 q=\pp{
2z_1+k_1/2&6z_4+3+3z_5\cr 
2z_2+1+k_2/2&3z_5\cr 
2z_3+k_3/2&6z_6+3z_5+1}.\ee
The number of $U(1)$s in the image of the extended lift, the
rank of q, is equal to one or two. Let us take the minimal
choice, rank one. Then the $k_j$ are even.
The cheapest
solution is \bb\tilde q=\pp{0&1\cr -1/2&0\cr
0&0},\qq q=\pp{0&3\cr 0&0\cr
0&1}\label{mscharge}\ee 
The image of the extended lift is
$\llll(U^n(\aa))=[U(1)\times SU(2)\times SU(3)]/[\zz_2
\times \zz_3]$ and  $\llll$ is precisely the fermionic group
representation of the standard model.

In section 2 the Higgs was obtained by fluctuating the
Dirac operator with all unitaries,
equation (\ref{spectralhiggs}). In the language of central
extensions this corresponds to including harmless $U(1)$s
and taking $\tilde q$ to be the $(M+N)\times (M+N)$ zero
matrix. This maximally extended lift is
single-valued, even and of rank three. If instead we take
the double-valued, even lift
$\llll$ of rank one defined by the charge matrix
(\ref{mscharge})
 and fluctuate the
Dirac operator with
$\llll(U^n(\aa))$ then we obtain only one doublet of Higgs
scalars and a fermionic mass matrix, that coincides with the
Dirac operator. In fact we reproduce Connes' version of the
standard model on the nose.

\subsection{Anomalies}

Since it coincides with the fermionic representation of the
standard model, the extended lift of (\ref{mscharge}) is free
of Yang-Mills and mixed gravitational-Yang-Mills anomalies
\cite{anom}. To spell out these conditions for a general
extended lift $\llll(w)$, equation (\ref{generallift}), we need
its infinitesimal version, $	\tilde \llll(X)$ defined as 
\bb\llll(1+X)=1+\tilde \llll(X)+O(X^2),
\qq X= (X_1,...,X_N) \in \bigoplus ^N_ {k=1}
u(n_k).\ee
The lift $\llll$ is free of Yang-Mills anomalies if
$\t[ \tilde \llll(X)^3\chi \varepsilon ]$ vanishes for all $X$
and it is free of mixed gravitational-Yang-Mills anomalies if
$\t[ \tilde \llll(X)\chi \varepsilon ]$ vanishes for all $X$. We
denote by $\varepsilon $ the projector on the particle space,
$\varepsilon ={\rm diagonal}\, (\rho _L(1),\rho
_R(1),0,0)$. For our model, $\llll$ is anomaly free if and only
if $q_{11}=3q_{31},\ q_{12}=3q_{32},\ q_{21}=q_{22}=0$.
In particular the lift is of rank one and its charge matrix can
be chosen to yield a double-valued, even lift,
\bb \tilde q=\pp{3z_3&6z_6+1\cr -1/2&0\cr z_3&2z_6
}.\label{free}\ee
We remark that the `bizzare' solution \cite{bizz} where only
the right-handed quarks have non-vanishing hypercharges
is not a central extension.

\section{So what?}

Since nine years Daniel Kastler asks the question whether the
unimodularity condition can be imposed before computing
the Higgs representation. In terms of spin lifts and
fluctuations of the metric, his question has a precise meaning
and a natural answer. Different central extensions do change
in general both the number of Yang-Mills bosons and the
number of Higgs bosons, while different unimodularity
conditions did not modify the number of Higgs bosons.
However in the standard model based on the algebra
$\cc\op\hhh\op M_3(\cc)$, the Higgs representation is the
same for all central extensions.

Recently Connes, Moscovici and Kreimer discovered a subtle
link between a noncommutative generalization of the index
theorem and perturbative quantum field theory. This link is a
Hopf algebra relevant to both theories \cite{cmk}. On the
other hand, Connes proposes to consider a Hopf algebraic
generalization of the spin group, in particular  quantum
$su(2)$ at third root of unity. Indeed this Hopf algebra
has 
$M_1(\cc)\oplus M_2(\cc)\oplus M_3(\cc)$ as quotient by
its nilradical \cite{tresch, daniel}.

\vskip1cm\noindent
It is a pleasure to acknowledge Jos\'e  Gracia-Bond\'\i a and
Bruno Iochum's friendly advice.

 \end{document}